\documentclass[]{aa}

\usepackage{pgf}
\usepackage{graphicx}
\usepackage{txfonts}
\usepackage{natbib,twoopt}
\usepackage{hyperref} 
\usepackage{caption} 
\usepackage{subcaption}
\bibpunct{(}{)}{;}{a}{}{,}             
\makeatletter
  \newcommandtwoopt{\citeads}[3][][]{\href{http://adsabs.harvard.edu/abs/#3}%
    {\def\hyper@linkstart##1##2{}%
     \let\hyper@linkend\@empty\citealp[#1][#2]{#3}}}
  \newcommandtwoopt{\citepads}[3][][]{\href{http://adsabs.harvard.edu/abs/#3}%
    {\def\hyper@linkstart##1##2{}%
     \let\hyper@linkend\@empty\citep[#1][#2]{#3}}}
  \newcommandtwoopt{\citetads}[3][][]{\href{http://adsabs.harvard.edu/abs/#3}%
    {\def\hyper@linkstart##1##2{}%
     \let\hyper@linkend\@empty\citet[#1][#2]{#3}}}
  \newcommandtwoopt{\citeyearads}[3][][]%
    {\href{http://adsabs.harvard.edu/abs/#3}
    {\def\hyper@linkstart##1##2{}%
     \let\hyper@linkend\@empty\citeyear[#1][#2]{#3}}}
\makeatother

\begin{document}

\title{Candidate star clusters toward the inner Milky Way 
discovered on deep-stacked $K_S$-band images from the VVV Survey}

\author{
Valentin D. Ivanov\inst{1,2}
\and
Andr\'es E. Piatti\inst{3,4}
\and
Juan-Carlos Beam\'in\inst{5,6}
\and
Dante Minniti\inst{7}
\and
Jordanka Borissova\inst{5,6}
\and
Radostin Kurtev\inst{5,6}
\and
Maren Hempel\inst{8}
\and
Roberto K. Saito\inst{9}
}

\offprints{V. Ivanov, \email{vivanov@eso.org}}

\institute{
European Southern Observatory, Ave. Alonso de C\'ordova 3107, 
Vitacura, Santiago, Chile
\and
European Southern Observatory, Karl-Schwarzschild-Str. 2, 
85748 Garching bei M\"unchen, Germany
\and 
Observatorio Astron\'omico, Universidad Nacional de C\'ordoba, 
Laprida 854, 5000, C\'ordoba, Argentina
\and
Consejo Nacional de Investigaciones Cient\'{\i}ficas y T\'ecnicas, 
Av. Rivadavia 1917, C1033AAJ, Buenos Aires, Argentina
\and
Instituto de F\'isica y Astronom\'ia, Facultad de Ciencias, Universida 
de Valpara\'iso, Av. Gran Breta\~na 1111, Valparaíso, Chile
\and
The Millennium Institute of Astrophysics, Santiago, Chile
\and
Departamento de Ciencias F\'isicas, Universidad Andr\'es Bello, 
Fern\'andes Concha 700, 759-1598 Las Condes, Santiago, Chile
\and
Instituto de Astrof\'isica, Pontificia Universidad Cat\o'lica de Chile, 
Av. Vicu\~na Mackenna 4860, Santiago, Chile
\and
Departamento de  F\'{i}sica, Universidade Federal de Santa Catarina, 
Trindade 88040-900, Florian\'opolis, SC, Brazil.
}

\date{Received 2 November 1002 / Accepted 7 January 3003}

\abstract 
{The census of star clusters in the inner Milky Way is incomplete
because of extinction and crowding.}
{We embarked on a program to expand the star cluster list in the 
direction of the inner Milky Way using deep stacks of K$_S$-band 
images from the VISTA Variables in Via L\,actea (VVV) Survey.}
{We applied an automated two-step procedure to the point-source 
catalog derived from the deep K$_S$ images: first, we identified
overdensities of stars, and then we selected only candidate 
clusters with probable member stars that match an isochrone with 
a certain age, distance, and extinction on the color-magnitude 
diagram.}
{This pilot project only investigates the cluster population in
part of one VVV tile, that is, b201. We identified nine cluster 
candidates and estimated their parameters. The new candidates are 
compact with a typical radius on the sky of $\sim$0.2--0.4\,arcmin
($\sim$0.4--1.6\,pc at their estimated distances). They are located 
at distances of $\sim$5--14\,kpc from the Sun and are subject to 
moderate extinction of E($B$$-$$V$)=0.4--1.0\,mag. They are sparse, 
probably evolved, with typical ages log($t$/1\,yr)$\sim$9. Based on 
the locations of the objects inside the Milky Way, we conclude that 
one of these objects is probably associated with the disk or halo 
and the remaining objects are associated with the bulge or the halo.
}
{The cluster candidates reported here push the VVV Survey cluster 
detection to the limit. These new objects demonstrate that the VVV 
survey has the potential to identify thousands of additional cluster 
candidates. The sub-arcsec angular resolution and the near-infrared 
wavelength regimen give it a critical advantage over other surveys.}

\authorrunning{V. Ivanov et al.}
\titlerunning{Candidate Star clusters toward the inner Milky Way}

\maketitle

\section{Introduction}\label{sec:intro}

The star cluster searches in the general direction of the inner 
Milky Way are hampered by two obstacles: extinction and crowding. 
The clumpy structure of the dust makes it even harder to find
clusters because both the often used visual inspection and the 
algorithms that identify density peaks can 
easily be deceived by holes in the dust or by sharp stellar density 
variation near the edges of dark clouds. It is not surprising that 
the first new generation cluster searches based on the Two Micron 
All Sky Survey \citep[2MASS;][]{2003yCat.7233....0S} were limited
to the vicinity of known objects that might be associated with 
clusters. 
For example, \citet{2000A&A...359L...9D} and \citet{2001A&A...376..434D} 
looked for clusters around known HII regions or unidentified IRAS 
sources and in known star-forming regions, respectively. 

\citet{2002A&A...394L...1I} and \citet{2003A&A...411...83B} 
attempted to find clusters blindly searching the 2MASS point source 
catalog for overdensities with automated tools and found some richer
clusters. However, subsequent works 
\citep[e.g.,][]{2005ApJ...635..560M,2006A&A...447..921K,2008A&A...486..771K,2015NewA...34...84C}
continued to discover more objects, even from the same observational 
data sets, indicating that the cluster census still remained 
incomplete. Indeed, incompleteness was detected even in the most 
comprehensive clusters catalogs to date by other teams; 
\citet{2014A&A...568A..51S} identified a lack of old (t$\geq$1\,Gyr) 
open clusters as close as the nearest 1\,kpc from the Sun in the 
list of nearly $\sim$3800 clusters reported by 
\citet{2013A&A...558A..53K}.

New deeper surveys with better angular resolution became available in 
the meantime, in particular the VISTA Variables in Via L\,actea 
\citep[VVV;][]{2010NewA...15..433M,2012A&A...537A.107S}. This motivated 
us to continue the search for new clusters. So far we have 
discovered in the VVV data in a total of 735 new star cluster candidates
\citep{2011A&A...527A..81M,2011A&A...535A..33M,2011A&A...532A.131B,
2014A&A...569A..24B,2014A&A...562A.115S,2015A&A...581A.120B}, greatly 
improving the census of the star clusters in the Galaxy. Most of the 
newly identified objects are practically invisible in the optical.

Here we describe the results from a pilot project to identify cluster 
candidates aimed (i) as a proof of concept and (ii) to evaluate, at 
least approximately, the expected number of candidates yielded from a 
full VVV search. This project is a precursor for the
LSST\footnote{\url{https://www.lsst.org/}} cluster searches, even 
though the LSST works in the optical, stacking multiple epochs as 
carried out in this work will generate deep images in the Milky Way. 
For simplicity, we ensured homogeneity by applying our algorithm to 
only one pawprint from one of the 348 VVV survey tiles, referred to 
in VVV as b201. There were no known clusters or cluster candidates in 
the surveyed area.
The explored $\sim$1$\times$1.5\,deg$^2$ area is centered at Galactic 
coordinates ($l$, $b$)$\sim$(350.5\,deg, $-$9.5\,deg) near the edge 
of the bulge. A tile near the outer edge of the VVV footprint was 
chosen for this test because it is less affected by extinction and 
crowding than the inner bulge and disk tiles.

\section{Observational data}\label{sec:data}

\subsection{VVV Survey}

The VVV is an European Southern Observatory (ESO) public survey of 
$\sim$562\,deg$^2$ of the Milky Way 
\citep{2010NewA...15..433M,2012A&A...537A.107S}, split between the 
bulge and the southern disk. The survey was completed in 2016.
The survey area was covered once, quasi-simultaneously in $ZYJHK_S$,
and then 60-100 $K_S$ band epochs were obtained over $\sim$5--6\,yr 
period for variability studies. The main goal of the VVV survey
is to map our galaxy with RR\,Lyr and Cepheids
in three dimensions \citep{2015ApJ...812L..29D,2016A&A...591A.145G}. 
But the enormous wealth of data generated by the VVV survey allows 
us to address a number of other questions, from proper motions 
\citep{2013A&A...557L...8B,2013A&A...560A..21I} to stellar clusters
\citep{2011A&A...532A.131B,2014A&A...569A..24B,2015A&A...581A.120B},
variable stars \citep{2016MNRAS.462.1180N} and even extragalactic 
sources in the zone of avoidance \citep{2014A&A...569A..49C}; this 
list is far from complete.

The survey was carried out with VISTA \citep[Visual and Infrared 
Survey Telescope for Astronomy;][]{2006Msngr.126...41E}, which is 
the ESO 4.1 m telescope located on Cerro Paranal. This telescope
is equipped with VIRCAM 
\citep[VISTA InfraRed CAMera;][]{2006SPIE.6269E..0XD}, which is a 
wide field near-infrared imager producing $\sim$1$\times$1.5\,deg$^2$ 
tiles\footnote{Tiles 
are contiguous images that combine six pawprints taken in an offset 
pattern; a pawprint is an individual VIRCAM pointing that generates a 
noncontiguous image of the sky because of the gaps between the 16 
detectors. \citet{2010NewA...15..433M} gives more details on the 
observing strategy of the VVV.}. The detectors are sensitive in the 
range from 0.9\,$\mu$m to 2.4\,$\mu$m. The data are processed with 
the VISTA Data Flow System 
\citep[VDFS;][]{2004SPIE.5493..411I,2004SPIE.5493..401E} pipeline 
at the Cambridge Astronomical Survey 
Unit\footnote{\url{http://casu.ast.cam.ac.uk/}} (CASU). 
The data products are available either from the ESO Science Archive 
or from the specialized VISTA Science 
Archive\footnote{\url{http://horus.roe.ac.uk/vsa/}}.
\citep[VSA;][]{2012A&A...548A.119C}.

\subsection{Deep-stacked K$_S$-band images}

Stacking $K_S$ images taken at multiple epochs allows us to obtain 
deeper data. We chose to combine only the best seeing 
($\leq$1.0\,arcsec) epoch available until January 2015. In the 
case of tile b201, this constituted 35 images. Pawprints rather than 
tiles were stacked together for two reasons: first, this ensures 
that images from the same detectors were combined with the same 
detector characteristics (read noise, gain, etc.); second, the 
sources are usually located in the same region of each detector, 
making the final point spread function more stable than what it 
would have been if images with large offsets were combined.

The stacking was performed with the {\it casutools} v1.0.30 task 
{\it imstack}\footnote{\url{http://casu.ast.cam.ac.uk/surveys-projects/software-release/stacking}}.
The task makes use of the WCS; when source CSAU catalogs are 
provided as input for each epoch, as in our case, it does fit the 
plate coefficients in pixel space to refine the WCS in the input 
header relative to the reference image.
Then, each image is resampled to the reference image, and the 
clipped average of each pixel scaled by the exposure time and 
weighted by the confidence value from each corresponding confidence 
map, is computed and recorded in the final image.

Next, we performed aperture photometry on the stacked image with 
the {\it casutools} task 
{\it imcore}\footnote{\url{http://casu.ast.cam.ac.uk/surveys-projects/software-release/imcore}}
with almost the same parameters as used by CASU to produce 
the single epoch source catalogs for VVV survey; we only adjusted 
the values of FWHM accordingly and set the radius of aperture for 
default photometric analysis ({\it rcore} parameter) to three 
pixels ($\sim$1\,arcsec). The final aperture radius used to 
compute the photometry is three times {\it rcore}. The output 
consists of fits images, confidence maps, and source catalogs 
with identical structure and content as their single-epoch 
counterpart produced by CASU for the individual epochs.

\begin{figure}
\begin{center}
\includegraphics[width=8.0cm]{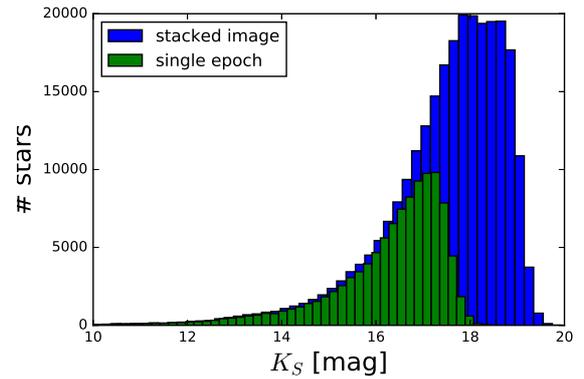} 
\end{center}
\caption{Number counts in single epoch (green) and in 
stacked image (blue) within 0.2\,mag wide bins. The stacking 
increases the number counts by a factor of $\geq$2.5 and the 
depth by nearly 2\,mag.}\label{fig:stack_comp1}
\end{figure}

The improvement in number counts, depth, and photometric errors is 
shown in Figs.\,\ref{fig:stack_comp1} and \ref{fig:stack_comp2}; 
the number of sources increased by a factor of $\sim$2.5 and the 
photometric error decreased considerably, particularly at the 
fainter end. We also cross-checked the photometry from the stacked 
image against the reference image (Fig.\,\ref{fig:phot_diff}); the 
median difference is 0.035\,mag with a median absolute deviation 
of 0.067\,mag, so the difference is consistent with zero.

\section{Cluster search and analysis}\label{sec:search}

We applied the same search procedure as in 
\citet{2016MNRAS.460..383P}, and we give here only a short 
summary of the steps. This procedure is based on identifying 
stellar surface 
density peaks after smoothing the surface density distribution 
with a kernel density estimator (KDE), which makes the density 
estimates independent from the bin size that is adopted in a 
``classical'' two-dimensional histogram. We used the Python 
implementation available form the {\sc 
ASTROMIL}\footnote{\url{http://www.astroml.org/index.html}}
library \citep{2012cidu.conf...47V}; {\it Gaussian} and {\it
tophat} KDEs were applied to the data with three different 
bandwidths of 0.23, 0.45, and 0.68 arcmin. In practical terms the
kernels replace the individual points/stars, and then they are 
added gather to create continuous and smooth surface density map. 

We run six different kernel overdensity searches on a sample of
$\sim$266,000 stars from one pawprint from the stacked deep 
$K_S$-band image. Similar to \citet{2016MNRAS.460..383P} we adopted a 
cutoff density of 0.05\,arcsec$^{-2}$, which is a factor of 1.25 
higher than the typical background surface density of 
0.04\,arcsec$^{-2}$. This selection yielded 323 cluster 
candidates. We performed a visual inspection of the stacked $K_S$ 
image and three-color $JHK_S$ image made from single-epoch images of 
the VVV survey (Fig.\,\ref{fig:finders}), and we discard any 
candidates that could not be associated with obvious, 
well-pronounced stellar cluster-like concentrations. We estimated 
during the inspection the sizes and central positions of the 
candidates (Table\,\ref{tab:candadates}). This left us 
with 36 candidates. Perhaps, some sparse clusters remained among 
the omitted object, but we prefer to err on the side of caution, 
instead of including questionable candidates in our sample.

\begin{figure}
\includegraphics[width=8.8cm]{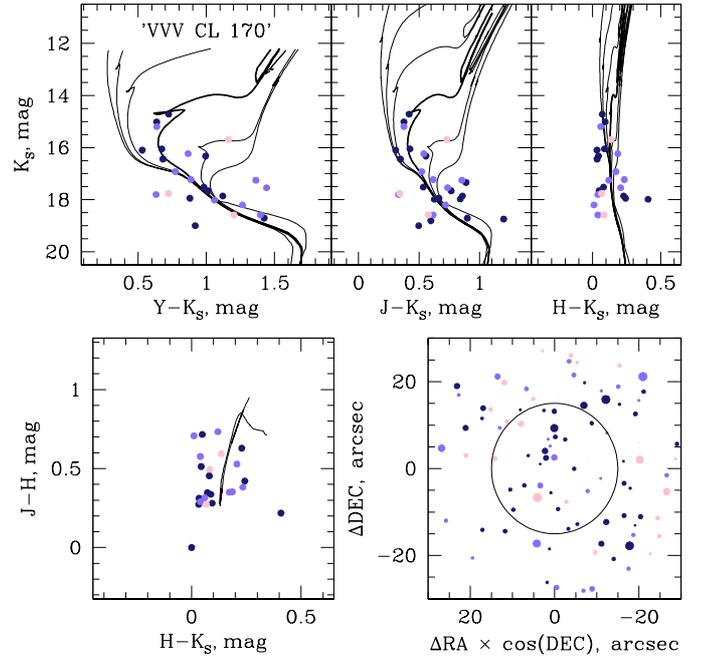}\\
\caption{
{\it Top:} Color-magnitude diagrams of the cluster VVV\,CC\,170 (top).
Solar abundance theoretical isochrones for ages log($t$/1\,yr) = 7.5, 
8.0, 8.5, 9.0, 9.5, and 10.0 from \citet{2012MNRAS.427..127B} are 
also shown.
{\it Bottom left:} Color-color diagram. Only the log($t$/1\,yr) = 9.0 
isochrone is drawn for simplicity. 
{\it Bottom right:} A map of the cluster candidate. The black 
circle indicates the object size. 
On all panels the cluster candidate membership probabilities are color 
coded: dark blue implies $\geq$75$\%$, light blue is 50\%, and pink 
indicates $\leq$25\%, respectively. Similar plots for the other cluster 
candidates are shown in Fig.\,\ref{fig:cmds_rest}.}\label{fig:cmds_1}
\end{figure}

\begin{figure}
\centering
\includegraphics[width=7.0cm]{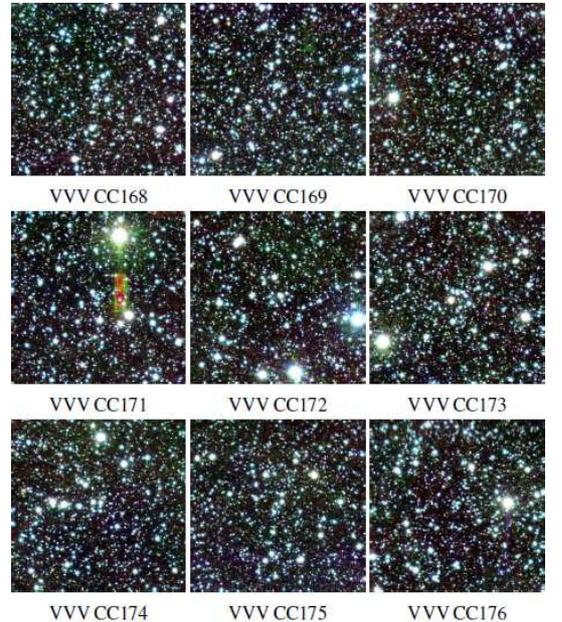}
\caption{Three-color 3$\times$3\,arcmin$^2$ finding charts 
($Y$ - blue, $J$ - green, $K_S$ - red) of the candidate 
clusters. North is at the top left and east is to the 
top right.}\label{fig:finders}
\end{figure}

The field star contamination was subtracted according the 
procedure of \citet{2012MNRAS.425.3085P}. First we selected four 
comparison fields, each one with the same area as the cluster 
candidate, in the vicinity of the cluster. Then for each field 
star on the $K_S$ versus $Y$$-$$K_S$ color-magnitude diagram (CMD), 
we removed on the CMD of the cluster the nearest star to that 
field star. In the process we counted how many times each cluster 
star remained in the cleaned sample: if the cluster star remained 
all four times, 
the star was assigned 100\% probability to be a member; if it 
remained three times, the membership probability was 75\%, and so 
on. In the final analysis we only considered stars that have 
$\geq$50\% likelihood of being members. An example of the results 
from the cleaning for the first object in our sample VVV\,CC\,170 
can be seen in Fig.\,\ref{fig:cmds_1}. Diagrams for the rest of the 
cluster candidates are shown in Fig.\,\ref{fig:cmds_rest}. The 
plots for a given cluster contains CMDs of the cluster for 
different colors (top), a color-color diagram (bottom left), and a 
map that shows the location of the cluster with a black circle 
(bottom right). The cluster membership probability of individual 
stars is color coded; pink, light blue, and dark blue represent 
$\leq$25\%, 50\%, and $\geq$75$\%$, respectively. 

Finally, we added to the CMDs theoretical solar-abundance isochrones 
for ages log($t$/1\,yr) = 7.5, 8.0, 8.5, 9.0, 9.5, and 10.0 from 
\citet{2012MNRAS.427..127B}. The isochrones were adjusted manually 
to reach the best fit to the data. The derived color excesses 
E($B$$-$$V$), distance moduli ($m$$-$$M$)$_0$, and ages log($t$/1\,yr) 
are listed in Table\,\ref{tab:candadates}. Conservatively, errors 
were tentatively adopted from the isochrones that bracket the best 
solution. Throughout this paper we use the extinction law of 
\citet{2006ApJ...638..839N,2008ApJ...680.1174N,2009ApJ...696.1407N}
and R$_V$=2.6 \citep{2013ApJ...769...88N}. We discarded many 
candidates during this step, typically because the scatter around 
the best-fitting isochrones were large, indicating differential 
reddening; this is expected if instead of a cluster there is a hole 
in the dust. The remaining nine candidates were listed in our final 
sample (Table\,\ref{tab:candadates}).

 
\begin{table*}[]
\caption{Final cluster candidate sample with positions and derived 
parameters: angular and linear radii, color excess E($B$$-$$V$), 
distance modulus (m$-$M)$_0$, heliocentric distance D, distance 
below the Galactic plane plane $-$Z, galactocentric distance 
R$_{GC}$, and age. See Sec.\,\ref{sec:search} for details. The 
cluster IDs continue the nomenclature last used in 
\cite{2014A&A...568A..16L}.}\label{tab:candadates}
\begin{center}
\begin{tabular}{cccccccccc}
\hline\hline
VVV & $\alpha$ $\delta$ (J2000) &Radius~&~E($B$$-$$V$)~&~~(m$-$M)$_0$~~& Distance D & $-$Z & R$_{GC}$ & ~Age~          \\
CC  & deg                       &arcmin (pc)~&~mag~    &~~mag~~        & kpc        & kpc  & kpc      & log($t$/1\,yr) \\
\hline
 168 & 18:00:50.2 $-$42:11:31 & 0.4  (1.3) & 1.0$\pm$0.5 & 15.2$\pm$0.5 & 11.0$_{-5.2}^{+ 9.9}$ & 1.8$_{-0.8}^{+ 1.6}$ & 3.8$_{-0.8}^{+1.6}$ & 9.0$\pm$0.5 \\   
 169 & 18:01:06.2 $-$42:04:15 & 0.25 (0.8) & 0.7$\pm$0.5 & 15.3$\pm$0.5 & 11.5$_{-5.4}^{+10.3}$ & 1.9$_{-0.9}^{+ 1.7}$ & 4.3$_{-0.8}^{+1.6}$ & 9.5$\pm$0.5 \\   
 170 & 18:01:03.0 $-$42:03:09 & 0.25 (0.6) & 1.0$\pm$0.5 & 14.5$\pm$0.5 &  7.9$_{-3.8}^{+ 7.1}$ & 1.3$_{-0.6}^{+ 1.2}$ & 1.9$_{-1.1}^{+1.5}$ & 9.0$\pm$0.5 \\   
 171 & 18:03:55.6 $-$42:20:07 & 0.25 (0.9) & 1.0$\pm$0.5 & 15.5$\pm$0.3 & 12.6$_{-5.8}^{+10.7}$ & 2.2$_{-1.0}^{+ 1.8}$ & 5.4$_{-0.7}^{+1.6}$ & 9.0$\pm$0.5 \\   
 172 & 18:06:06.6 $-$42:09:46 & 0.3  (0.6) & 0.5$\pm$0.5 & 14.2$\pm$0.5 &  6.9$_{-3.3}^{+ 6.2}$ & 1.2$_{-0.6}^{+ 1.1}$ & 2.0$_{-0.7}^{+1.2}$ & 9.5$\pm$1.0 \\   
 173 & 18:04:29.0 $-$41:41:33 & 0.3  (1.6) & 0.6$\pm$0.5 & 13.5$\pm$0.5 &  5.0$_{-2.7}^{+ 5.7}$ & 0.8$_{-0.4}^{+ 1.0}$ & 3.3$_{-0.0}^{+1.1}$ & 9.0$\pm$1.0 \\   
 174 & 18:02:34.6 $-$41:31:08 & 0.4  (0.4) & 0.9$\pm$0.5 & 15.7$\pm$1.0 & 13.8$_{-6.5}^{+12.4}$ & 2.2$_{-1.1}^{+ 2.0}$ & 6.4$_{-0.5}^{+1.1}$ & 9.0$\pm$1.0 \\   
 175 & 18:05:38.2 $-$41:43:58 & 0.3  (0.8) & 0.4$\pm$0.5 & 14.7$\pm$0.5 &  8.7$_{-4.1}^{+ 7.8}$ & 1.5$_{-0.7}^{+ 1.3}$ & 2.1$_{-1.0}^{+1.6}$ & 9.5$\pm$0.5 \\   
 176 & 18:05:00.8 $-$41:22:49 & 0.2  (0.7) & 0.9$\pm$0.3 & 15.5$\pm$0.5 & 12.6$_{-4.4}^{+ 6.7}$ & 2.1$_{-0.7}^{+ 1.1}$ & 5.3$_{-0.5}^{+0.9}$ & 9.0$\pm$0.5 \\   
\hline
\end{tabular}
\end{center}
\end{table*}

\section{Results and discussion}\label{sec:results}

\begin{figure}
\begin{center}
\includegraphics[height=6.0cm,width=6.0cm]{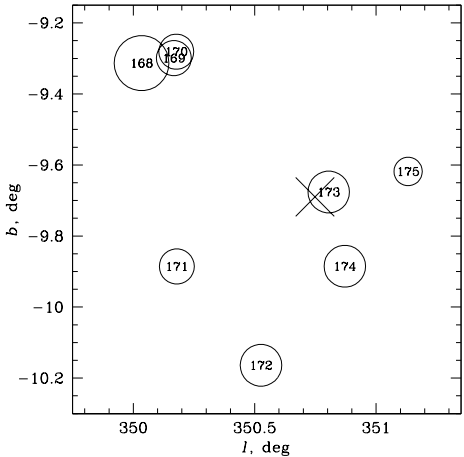}
\end{center}
\caption{Location of our new cluster candidates in Galactic 
coordinates. The circle size is linearly proportional to the 
cluster radius but out of scale; the radius spans the range 
0.2--0.4\,arcmin. The objects are numbered according to 
Table\,\ref{sec:search}. The x marks the center of tile 
b201.}\label{fig:map}
\end{figure}

The final list of identified cluster candidates contains nine 
objects. Their locations on the sky are shown in Fig.\,\ref{fig:map} 
and there appears to be some clustering: VVV\,CC\,169 and 
VVV\,CC\,170 have projected on-sky separation of $\sim$2\,arcmin, 
but these objects have different extinctions and distances 
(Table\,\ref{tab:candadates}), so it is unlikely that they are 
physically connected. A physical connection between VVV\,CC\,168 
and VVV\,CC\,169, separated by $\sim$9\,arcmin and located at 
similar distances is more likely, but given the uncertainties, 
can only be tentative. The derived distances range between 
$\sim$5\,kpc and $\sim$14\,kpc, and our candidates are located 
$\sim$1--2\,kpc below the Galactic plane. The candidate closest 
to us, VVV\,CC\,169, may be associated with the Milky Way disk 
or halo. The remaining objects are probably part of the Milky Way 
bulge or halo, which comes as no surprise because their CMDs are 
consistent with an evolved stellar population with 
log($t$/1\,yr)$\sim$9. The large uncertainties in the locations 
of the candidates and the lack of radial velocities prevent us 
from drawing firmer conclusions about which Milky Way component 
they belong to.

We attempted to verify the stellar nature of the probable members 
of our final cluster candidate from the morphological {\it class}
parameter\footnote{\url{http://horus.roe.ac.uk/vsa/www/vsa_browser.html}}
derived by the CASU pipeline for the deep tiles. Unfortunately, 
this parameter is heavily affected by the crowding in the field; 
for the investigated pawprint, the total percentage of extended 
sources is unrealistically high, at nearly 40\,\%, and we can not 
attribute this to background galaxies. Instead, we compared the 
apparent $Y$$-$$K_S$ color of the members with those 
for some confirmed background galaxies in the VVV Survey from 
\citet{2014A&A...569A..49C} and found virtually no overlap: the 
galaxies span $Y$$-$$K_S$$\sim$1.5--2.2\,mag, while the 
high-probability cluster members are bluer with 
$Y$$-$$K_S$$\leq$1.2--1.5\,mag (Figs.\,\ref{fig:cmds_1} and 
\ref{fig:cmds_rest}).

Next, we investigated whether the cluster candidates could be holes in 
the Milky Way dust through which we can see the bulge population. 
The new objects are compact, and most of them have radii 
15--18\,arcsec, which is smaller than the spatial resolution of 
even the finest available reddening maps of the inner Milky Way, 
for example, 1$\arcmin$$\times$1$\arcmin$ for \citet{2013A&A...552A.110G}.
This map reports for the fields of all our candidates 
E($J$$-$$K_S$)=0.0$\pm$0.1\,mag, which in E($B$$-$$V$) implies 
1$\sigma$ limits of $\sim$0.34\,mag, which agrees at 2--3$\sigma$
level with all our CMD based estimates. We also inspected the WISE 
\citep[Wide-field Infrared Survey Explorer;][]{2010AJ....140.1868W}
images and identified no obvious signature of circular dust 
emission around the objects.

The compact appearance of the new cluster candidates is probably 
the reason why they were not identified earlier 
\citet[e.g.,][]{2015NewA...34...84C}. In most cases the brightest 
member stars are in the range $K_S$$\sim$14--16\,mag, placing 
them near the confusion limit of 2MASS in the inner Milky Way or 
even below that. The fainter member stars are beyond 2MASS 
detection, which might be the reason why these objects were not 
identified in the 2MASS searches either. In some cases the few 
brightest member stars can be seen on the DSS2 red and DSS2 IR 
images, but without the fainter stars it is difficult to realize
that these may be overdensities associated with star clusters.

We treat the derived ages, 
extinctions, and distance moduli with care because an inspection 
of the CMDs in Figs.\,\ref{fig:cmds_1} and \ref{fig:cmds_rest} 
shows that most color excesses and distances can easily vary by 
$\sim$0.5\,mag, leading to a change of $\sim$0.5--1 in log($t$/1\,yr).

None of the new candidates appear to be extremely young and the
WISE images did not reveal any bright mid-infrared sources that 
may be associated with the new objects. The missing globulars 
that \citet{2005A&A...442..195I} predicted might hide among these 
candidates. Unfortunately, the extinction and the distance make 
it difficult to characterize these objects spectroscopically
\citet[e.g.,][]{2011MNRAS.416..465L} and further investigations 
might have to rely on photometric techniques 
\citet[e.g.,][]{2002A&A...390..937I,2009A&A...508.1279B}.

Our candidates have smaller apparent sizes (listed in 
Table\,\ref{tab:candadates}) than average radii 
$<$r$>$=0.6$\pm$0.4\,arcmin of the infrared-selected star 
clusters and stellar groups listed by \citep{2003A&A...397..177B}.
However, the physical radii of the two samples our objects are 
similar: the peak of the distribution of linear diameters 
\citep[figure 3 in ][]{2003A&A...397..177B} corresponds to radii 
of 0.5--1\,pc.
A comparison with older objects, 141 globular clusters from the 
list of \citet{1996AJ....112.1487H}, indicates that the radii 
of most our candidates are bracketed between the average core 
radius $<$R$_C$$>$=0.5$\pm$0.6\,pc and the average half-light 
radius $<$R$_H$$>$=1.3$\pm$0.8\,pc of the globulars.
Therefore, in terms of sizes the new candidates blend well with 
the samples of known cluster.

We conclude that the sub-arcsec angular resolution of the VVV 
survey in $K_S$ band was critical for uncovering these new cluster
candidates. This survey fills in a niche between optical surveys 
that cannot penetrate the dust and the mid-infrared WISE survey, 
which can penetrate the dust, but the WISE suffers from poor 
angular resolution. Our result hints that the GLIMPSE survey 
\citep[Galactic Legacy Infrared Mid-Plane Survey 
Extraordinaire;][]{2003PASP..115..953B}
and its various extensions may be a fertile ground for future 
searches because they combine the dust penetration of the 
mid-infrared with the high angular resolution of VVV.

Our selection procedure relies on an automated repeatable density 
peaks selection, but it also includes manual verification steps, so 
it is not truly objective. This makes a cluster candidate catalog 
derived following this technique unsuitable for statistical studies 
of {the clusters' spatial distribution}, age distribution, etc. If 
we simply scale up the number of candidates recovered from a single 
pawprint of a single VVV tile and keeping in mind that a pawprint 
only covers about 40\,\% of a tile, we can expect to recover 
thousands of candidates -- preferentially low luminosity objects -- 
from all 348 VVV tiles; the exact number is highly uncertain, given 
the limited number of the objects, but falls in line with the latest 
$\lambda$-CDM simulations \citep[e.g.,][]{2015MNRAS.446..521S}. This
suggests that there should be thousands of faint Milky Way building 
blocks and the cluster candidates that we find might be the remnants 
of those.

\section{Summary}\label{sec:summary}

We report results form an automated cluster selection based on 
deep stacked $K_S$-band images form the VVV survey. We identified 
more than 300 density peaks, but follow-up analysis 
indicated that the majority of them are likely false positives, so 
our final list contains only nine candidates. Most of these objects are 
probably evolved and many are associated with the bulge. The 
result from this search indicates that the entire VVV survey can 
yield thousands more Milky Way cluster candidates.

\begin{acknowledgements}
This paper is based on observations made with ESO telescopes at the La 
Silla Paranal Observatory under program ID 092.B-0104(A). We have made 
extensive use of the SIMBAD Database at CDS (Centre de Donn\'ees 
astronomiques) Strasbourg, the NASA/IPAC Extragalactic Database (NED), 
which is operated by the Jet Propulsion Laboratory, CalTech, under 
contract with NASA, and of the VizieR catalog access tool, CDS, 
Strasbourg, France. 
Support for JB, DM, JCB, RK, MH is provided by the Ministry of Economy, 
Development, and Tourism’s Millennium Science Initiative through grant 
IC120009, awarded to The Millennium Institute of Astrophysics, MAS.
Support for RK is provided from Fondecyt Reg. No. 1130140.
R.K.S. acknowledges support from CNPq/Brazil through project 
310636/2013-2.
We are grateful to the anonymous referee for useful suggestions 
that helped to improve the paper.
\end{acknowledgements}

\bibliographystyle{aa}
\bibliography{vvv_cl_07}

\begin{appendix}
\section{Photometric properties of the deep-stacked VVV images}

\begin{figure}[b]
\begin{center}
\includegraphics[width=8.0cm]{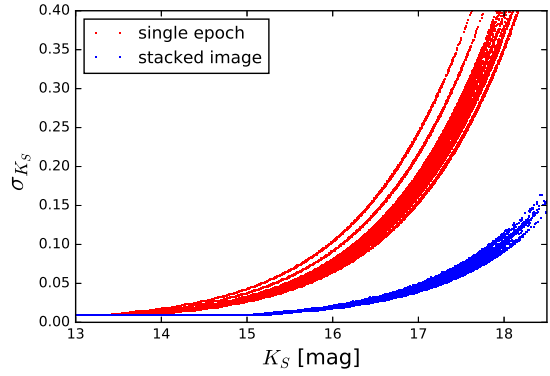} 
\end{center}
\caption{Photometric error as a function of $K_S$ magnitude 
for single epoch (red) and for a stacked image (blue). Only 
sources appearing in both catalogs are 
plotted.}\label{fig:stack_comp2}
\end{figure}

\begin{figure}[b]
\begin{center}
\includegraphics[width=8.0cm]{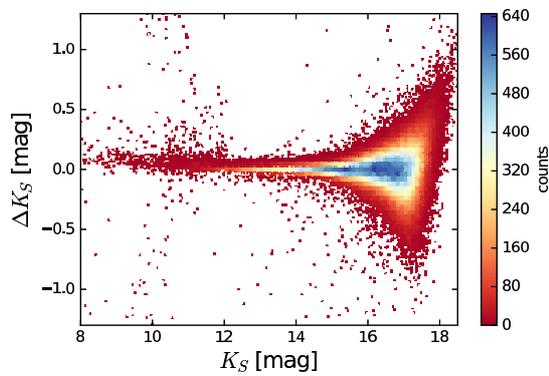}
\end{center}
\caption{Photometric difference between the sources in the reference 
image and the stacked image. There is a small offset in the median 
value, but this offset is consistent with zero given the scatter in 
the data. This difference is probably related to the lower signal to 
noise in the reference image.}\label{fig:phot_diff}
\end{figure}

\section{Color-magnitude diagrams}
\begin{figure*}[b]
\includegraphics[width=8.8cm]{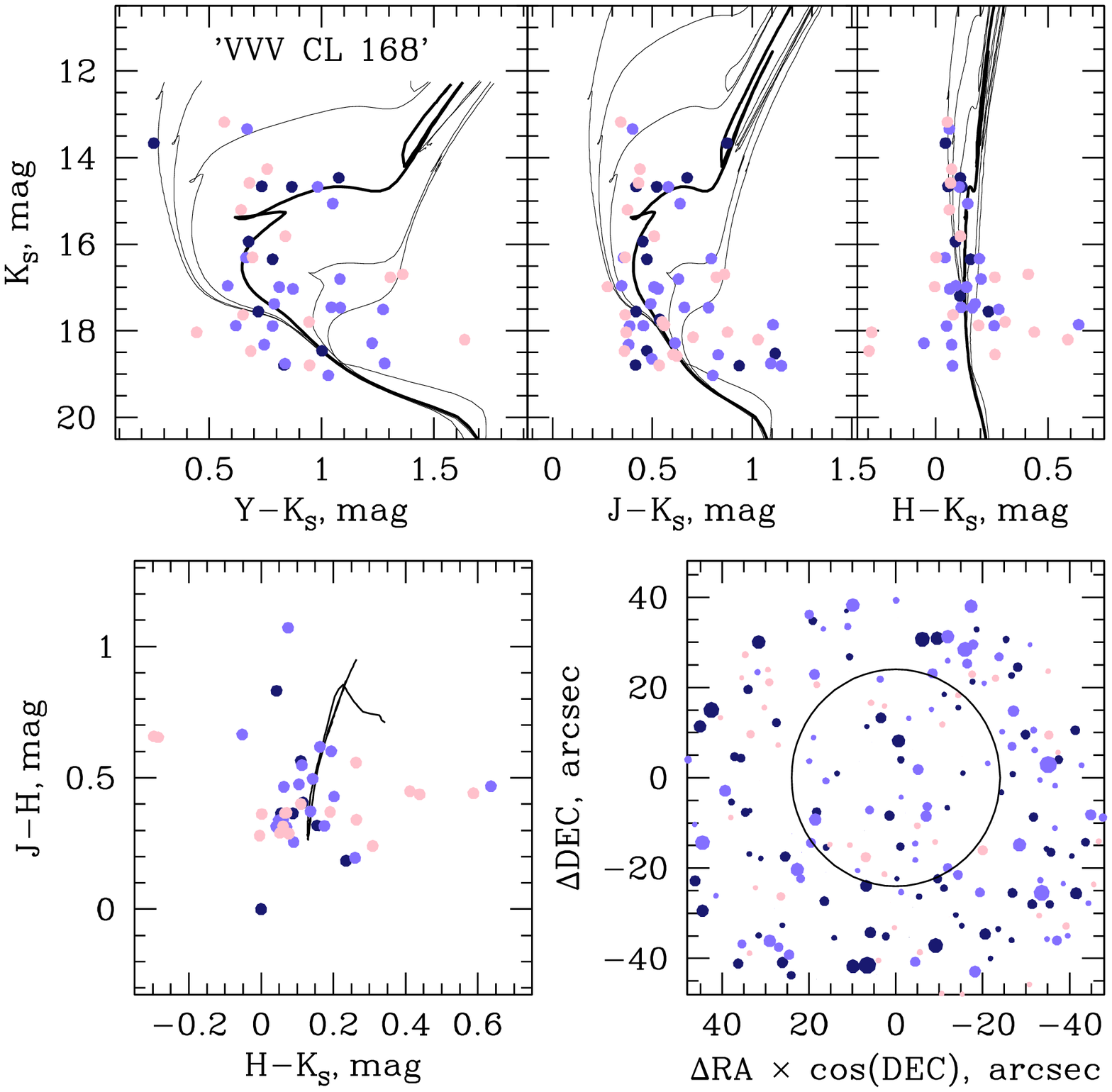} \includegraphics[width=8.8cm]{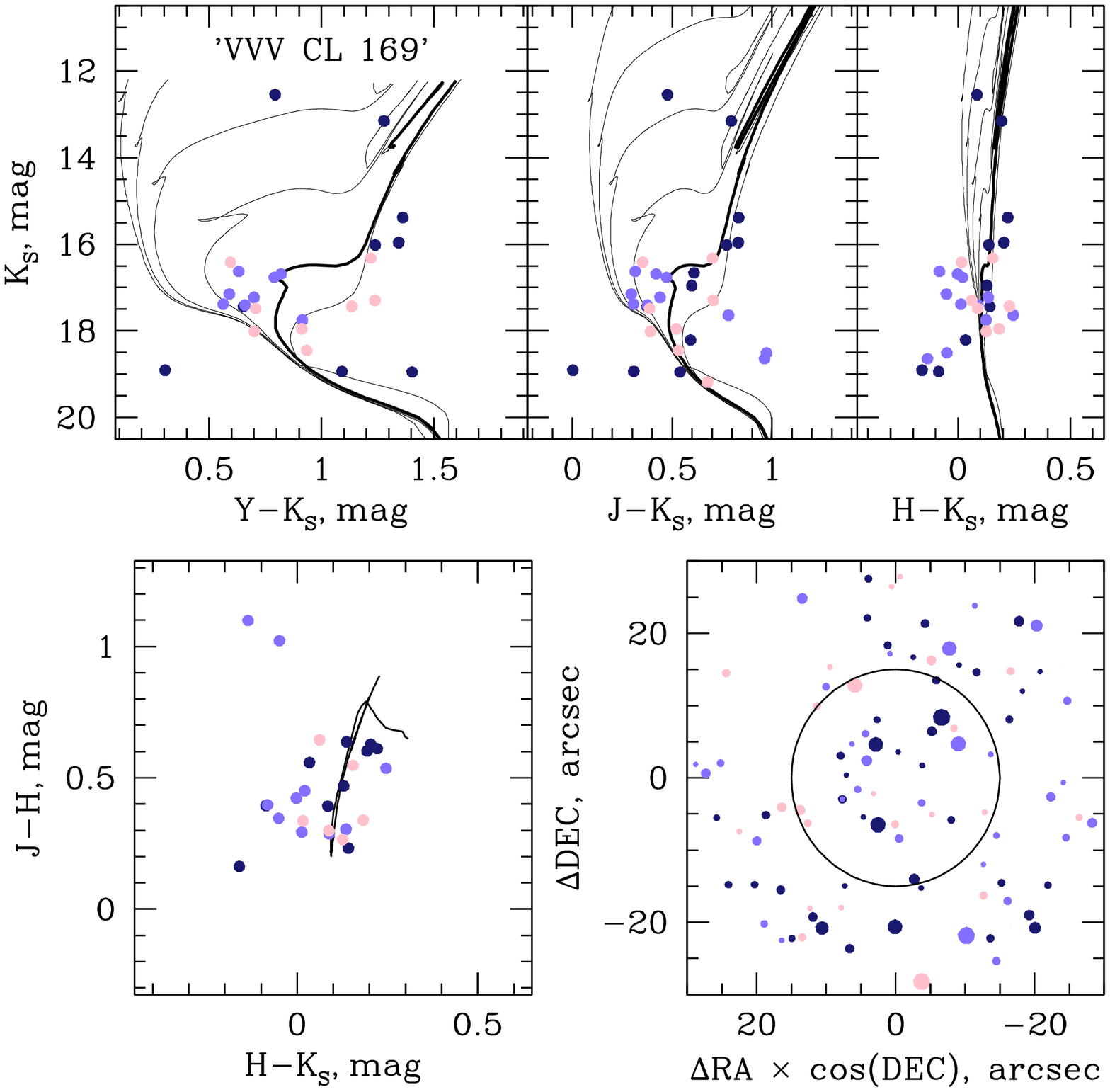} \\ 
\includegraphics[width=8.8cm]{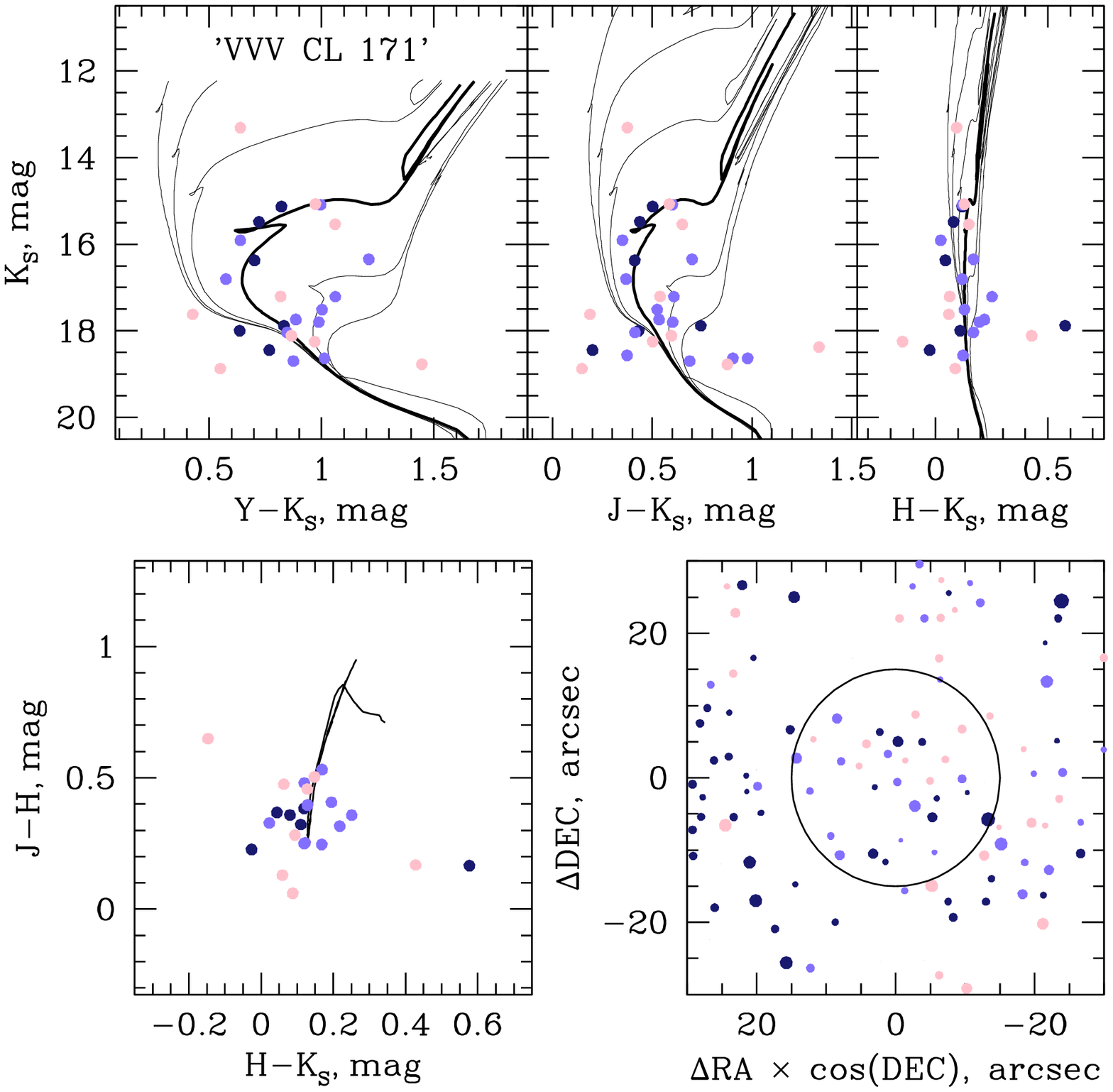} \includegraphics[width=8.8cm]{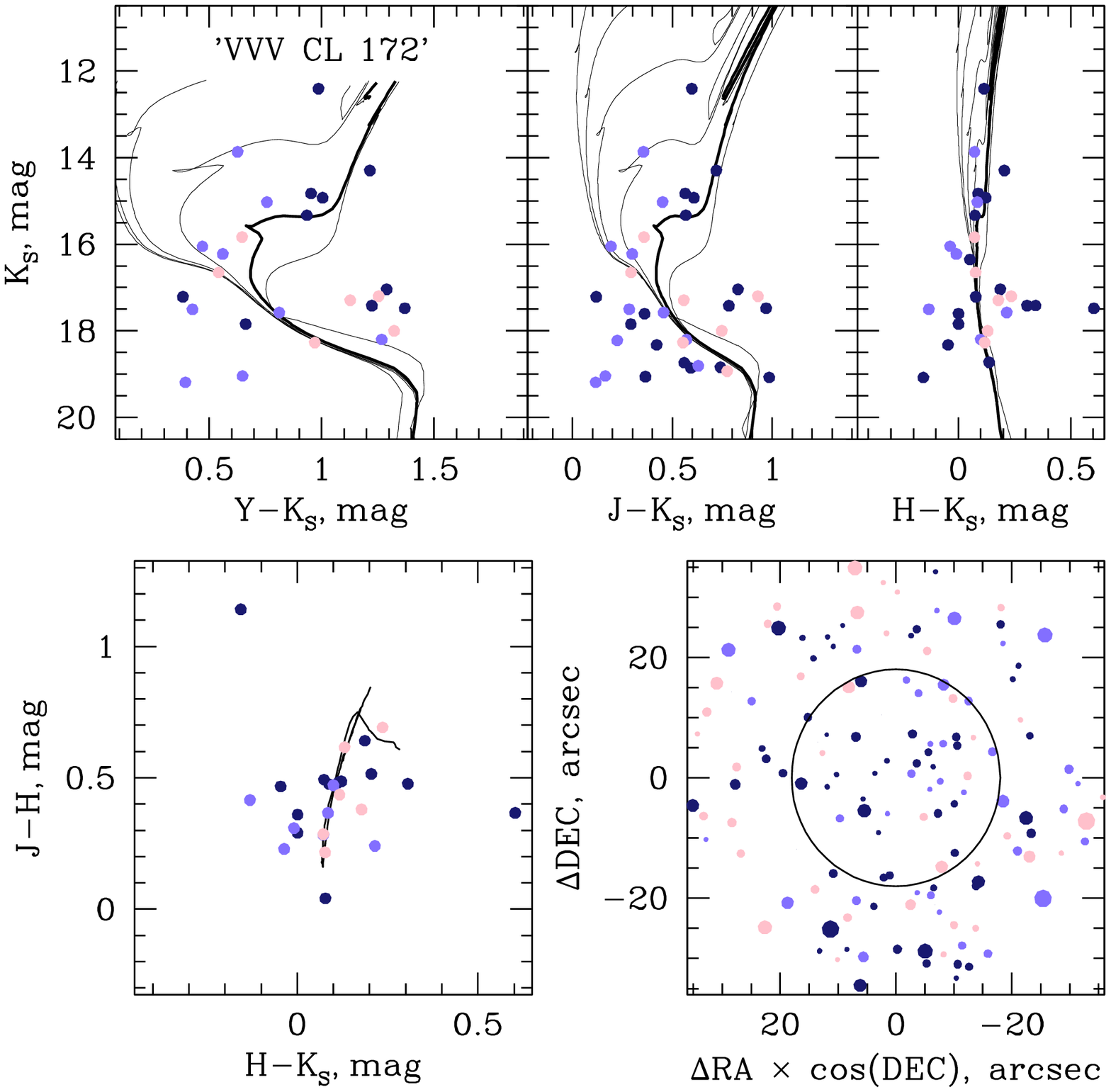} \\
\caption{Color-magnitude diagrams of the cluster candidates VVV\,CC\,168 
(top left), VVV\,CC\,169 (top right), VVV\,CC\,171 (bottom left), and 
VVV\,CC\,172 (bottom right). The subpanels and symbols are the same 
as in Fig.\,\ref{fig:cmds_1}}\label{fig:cmds_rest}
\end{figure*} 

\addtocounter{figure}{-1}
\begin{figure*}[b]
\includegraphics[width=8.8cm]{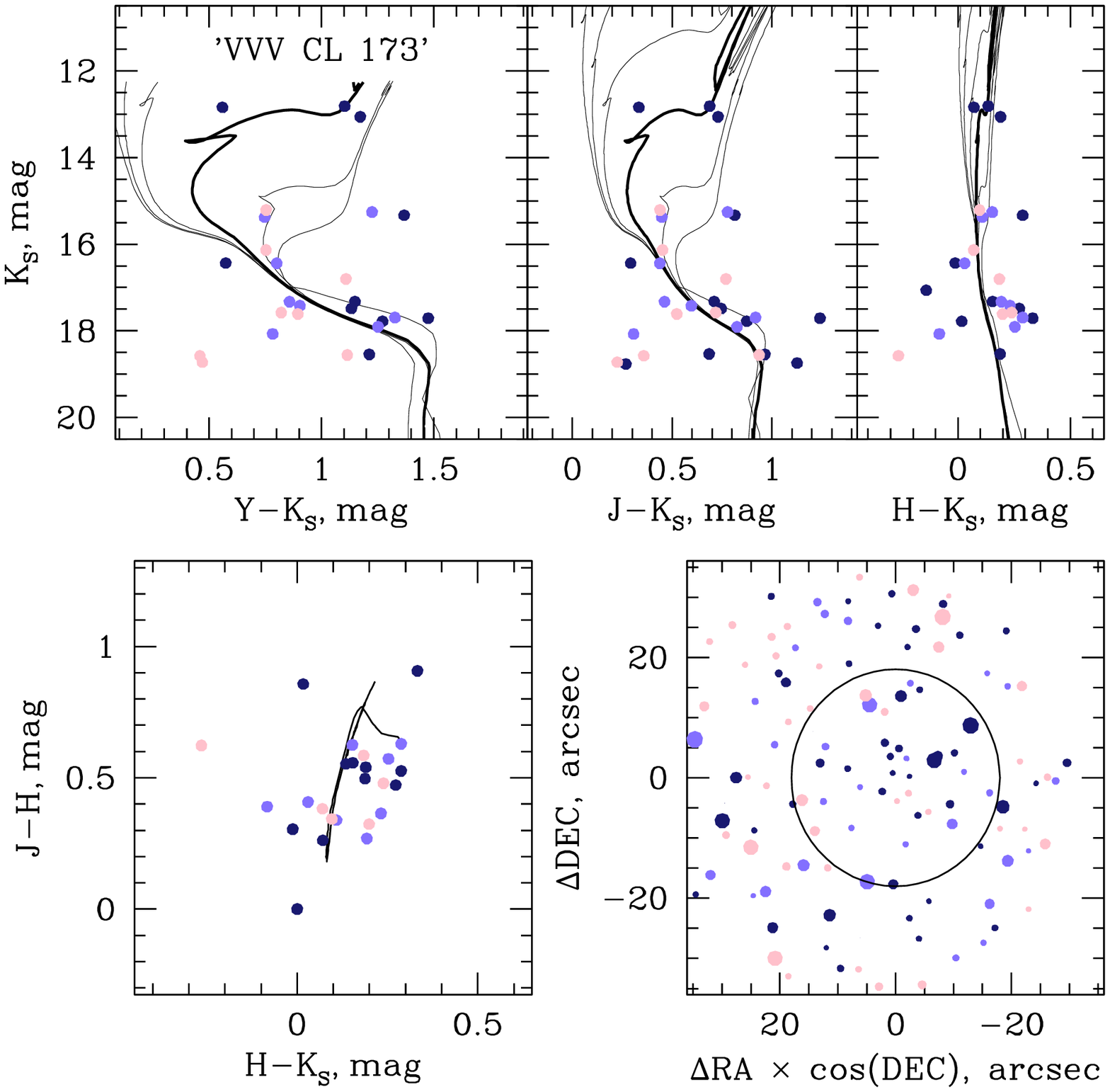} \includegraphics[width=8.8cm]{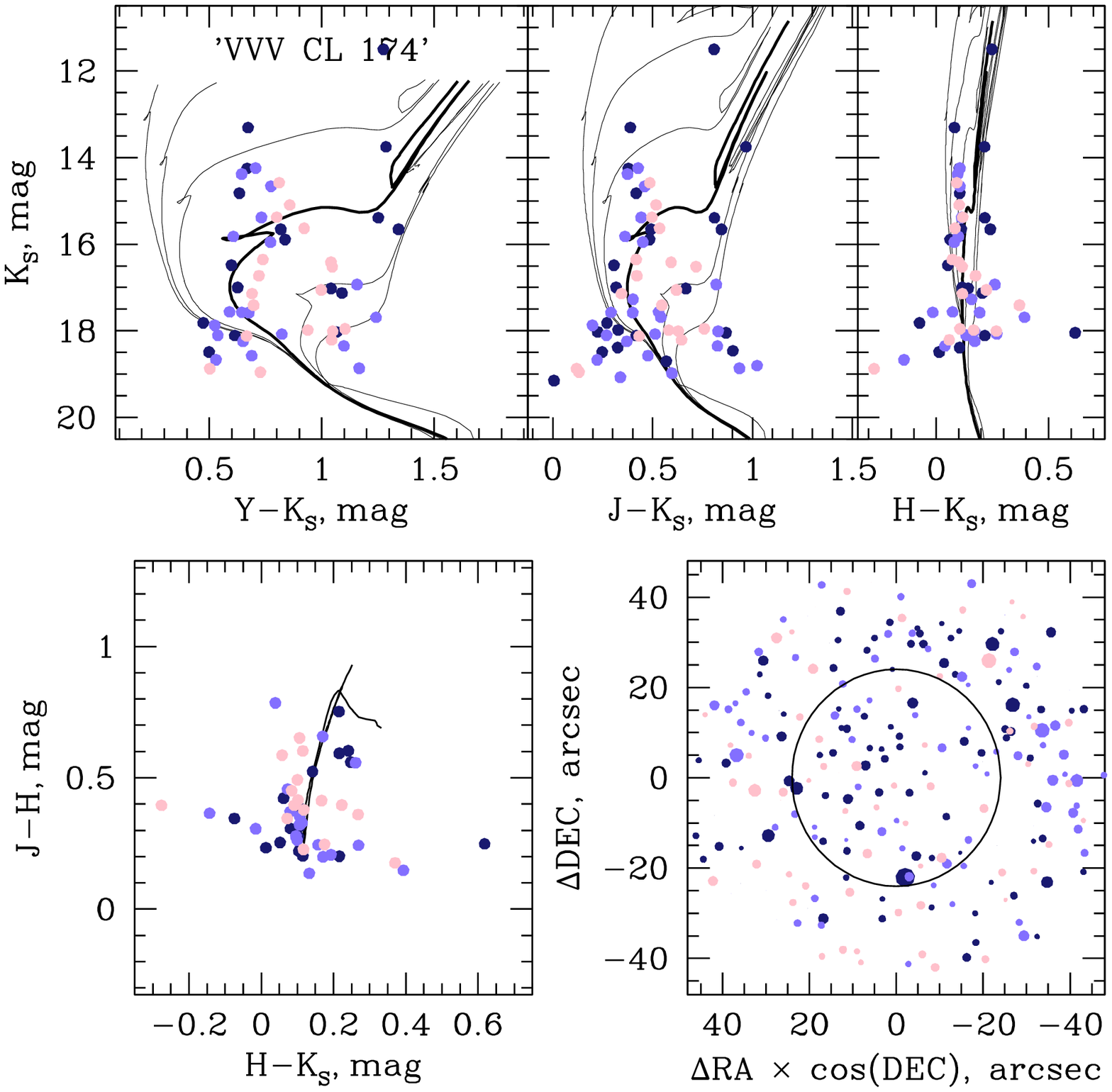} \\
\includegraphics[width=8.8cm]{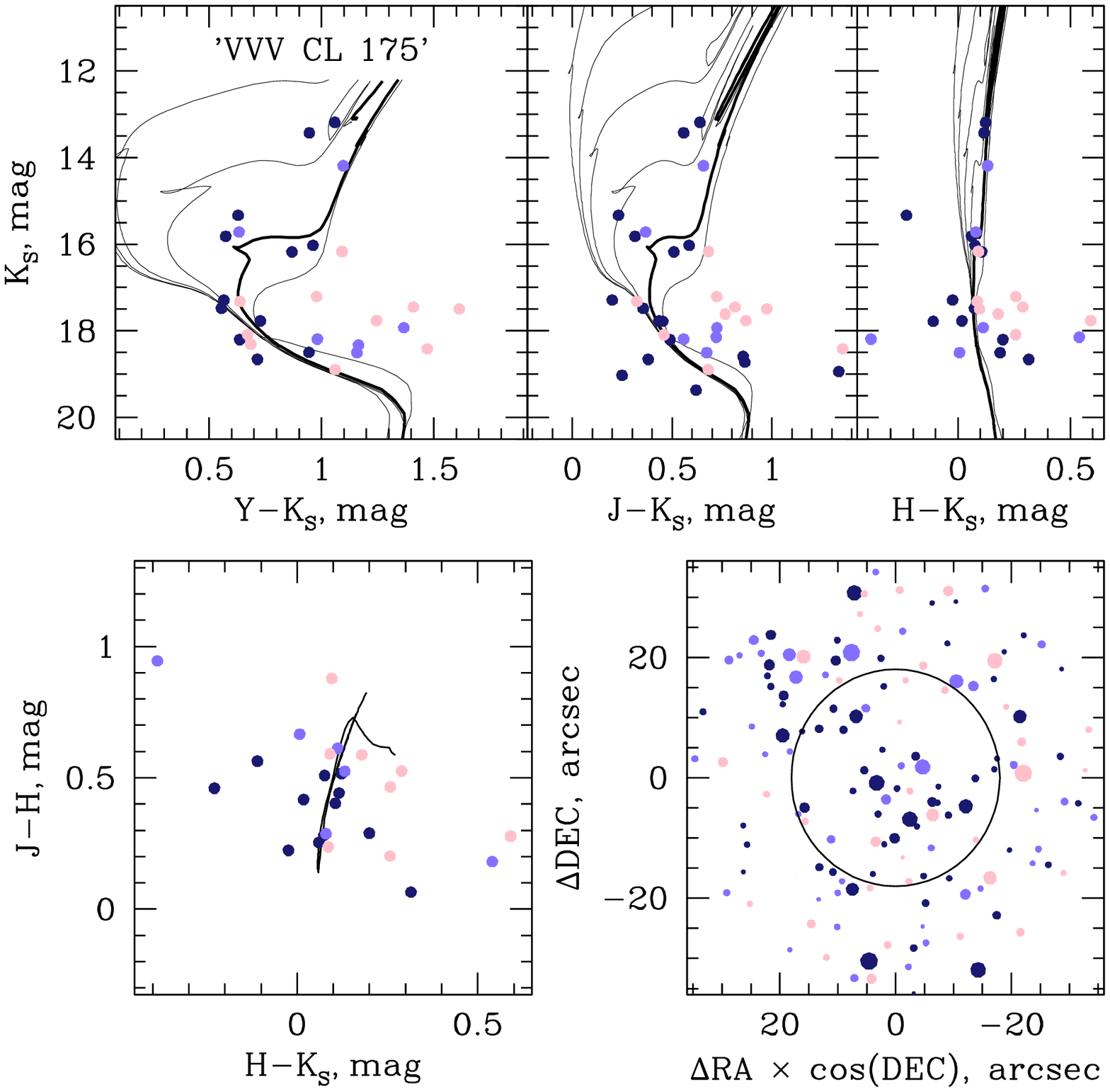} \includegraphics[width=8.8cm]{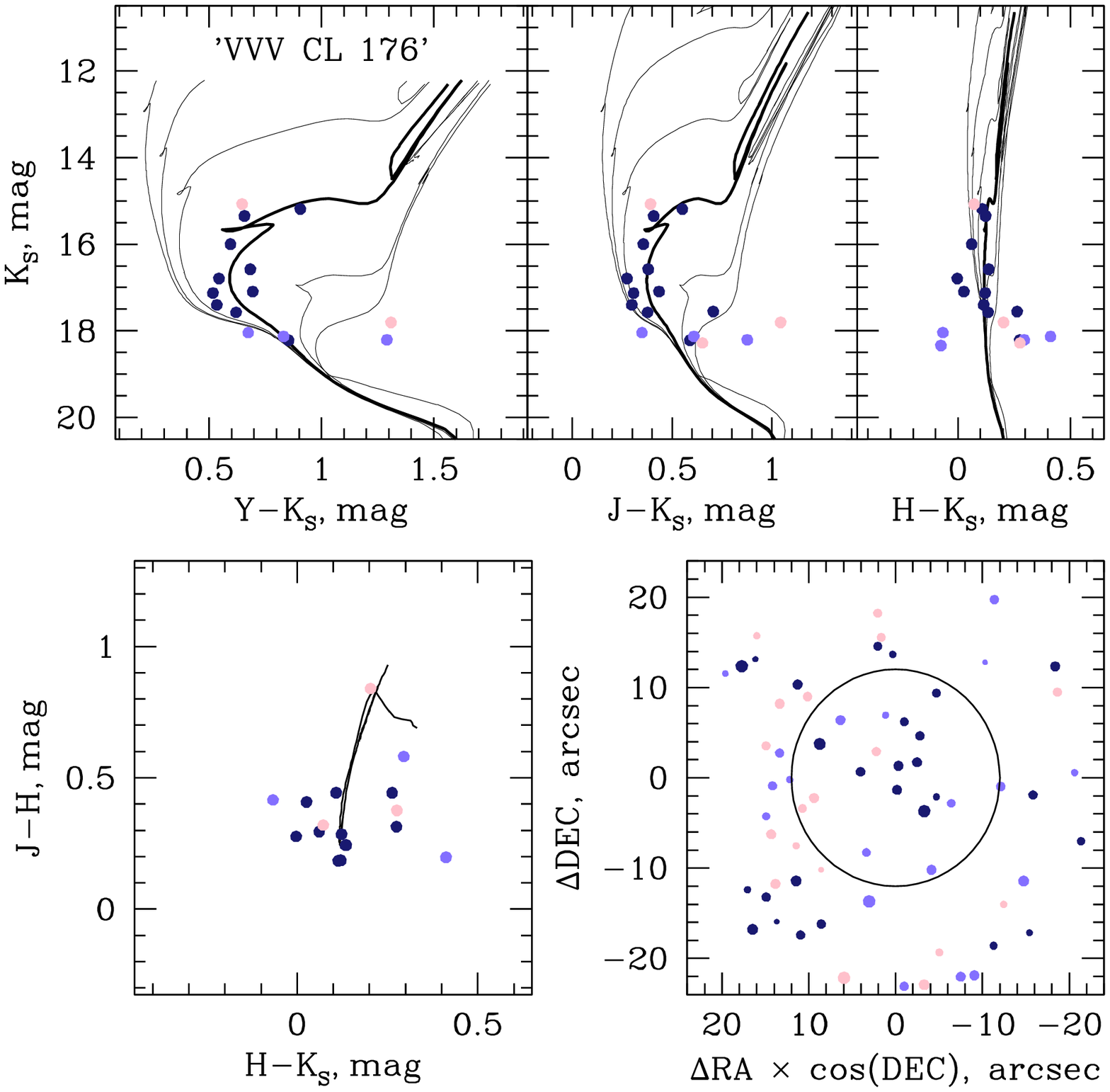} \\
\caption{Continued. Color-magnitude diagrams of the cluster candidates 
VVV\,CC\,173 (top left), VVV\,CC\,174 (top right), VVV\,CC\,175 (bottom 
left) and VVV\,CC\,176 (bottom right). The sub-panels and symbols are 
the same as in Fig.\,\ref{fig:cmds_1}}
\end{figure*} 
\end{appendix}

\end{document}